\documentclass[fleqn,10pt]{wlscirep}
\usepackage{graphicx}
\usepackage{epstopdf}
\usepackage{float}
\usepackage{color}
\usepackage{subcaption}

\title{Oscillating magnetoresistance due to fragile spin structure in metallic GdPd$_3$}

\author[1,2,3*]{Abhishek Pandey} 
\author[3]{Chandan Mazumdar}
\author[3]{R. Ranganathan}
\author[2]{D. C. Johnston}
\affil[1]{Department of Physics and Astronomy, Louisiana State University, Baton Rouge, Louisiana 70803, USA}
\affil[2]{Ames Laboratory-USDOE and Department of Physics and Astronomy, Iowa State University, Ames, Iowa 50011, USA}
\affil[3]{Experimental Condensed Matter Physics Division, Saha Institute of Nuclear Physics, 1/AF, Bidhannagar, Kolkata 700064, India}
\affil[*]{abhishek.phy@gmail.com}

\begin{abstract}
Studies on the phenomenon of magnetoresistance (MR) have produced intriguing and application-oriented outcomes for decades--colossal MR, giant MR and recently discovered extremely large MR of millions of percents in semimetals can be taken as examples. We report here the investigation of oscillating MR in a cubic intermetallic compound GdPd$_3$, which is the only compound that exhibits MR oscillations between positive and negative values.  Our study shows that a very strong correlation between magnetic, electrical and magnetotransport properties is present in this compound. The magnetic structure in GdPd$_3$ is highly fragile since applied magnetic fields of moderate strength significantly alter the spin arrangement within the system--a behavior that manifests itself in the oscillating MR. Intriguing  magnetotransport characteristics of GdPd$_3$ are appealing for field-sensitive device applications, especially if the MR oscillation could materialize at higher temperature by manipulating the magnetic interaction through perturbations caused by chemical substitutions. 
\end{abstract}

\begin{document}

\flushbottom
\maketitle

\thispagestyle{empty}

\section*{Introduction}
Investigation of the phenomenon of magnetoresistance (MR) has been of the central interest of the condensed matter physics, materials science and electrical and electronics engineering communities for decades. Materials that exhibit large MR as well as the physical and chemical properties that are optimum for applications are often used in devices, such as sensors and magnetic memory drives \cite{Moritomo-1996, Husmann-2002, Lenz-1990}. The discoveries of colossal MR \cite {Salamon-2001, Ramirez-1997} and giant MR  \cite{Baibich-1988, Binasch-1989} were very significant stepping stones in advancement of the field of MR studies and their applications.  Recently, the interest in the field was renewed after the discovery of extremely large positive MR (XMR) in nonmagnetic Weyl, Dirac, and resonant compensated semimetals and topological insulators \cite{Ali-2014, Tafti-2016, Liang-2015, Wang-2014, Wang-2015, Keum-2015, Shekhar-2015, Qu-2010, Wang-2012}. 

There are many reports on the experimental observations of MR oscillations within the positive MR regime mostly due to quantum effects, for example in GaAs/AlGaAs hetrostructures \cite{Gerhardts-1989, Mamani-2009}, black phosphorus quantum wells \cite{Tayari-2015}, and in nano systems {\it e.g.}, single-crystal nanobelts \cite{Wang-2009}, indium-oxide nanowires \cite{Johansson-2005}, niobium-nitride nanowires \cite{Patel-2009} and nanopatterned superconducting films \cite{Sochnikov-2010}. However, to our knowledge, MR oscillation between positive and negative values has not been reported for any magnetic compound except the cubic binary compound GdPd$_3$ (ref.~\citenum{Kitada-2011}). An unusual MR behavior was earlier reported in $Ln_2$Ni$_3$Si$_5$ ($Ln$ = Pr, Dy, Ho) compounds \cite{Mazumdar-1996}. The MR of these compounds shows only one small positive peak followed by a negative minimum. The three distinct crossovers between positive and negative values of MR observed in GdPd$_3$ are absent in $Ln_2$Ni$_3$Si$_5$ compounds. 

The $M$Pd$_3$ ($M$: Y and rare earth) compounds crystallize in the cubic AuCu$_3$ type  structure (space group: $Pm\bar3m$) \cite{Gardner-1972}. All the $M$Pd$_3$ compounds are metallic and depending upon the type of $M$ ion exhibit a variety of magnetic ground states \cite{Gardner-1972}. One member of the series, GdPd$_3$, exhibits antiferromagnetic (AFM) ordering below the N\'eel temperature $T_{\rm N} \sim 6$~K (refs.~\citenum{Gardner-1972, Pandey-2009a}). The value of $\xi = \chi(0)/\chi(T_{\rm N}) = 0.81$  ($\chi$: magnetic susceptibility) for polycrystalline GdPd$_3$ at applied magnetic field $H = 0.1$~T suggests a noncollinear AFM spin arrangement of the Gd spins where the ordered moments below $T_{\rm N}$ are not aligned along the same axis, as a collinear AFM structure would have otherwise resulted in $\xi = 2/3$ (refs.~\citenum {Johnston-2012, Johnston-2015}). 

In the present work, we investigate the low-temperature MR characteristics of GdPd$_3$ down to $T = 0.7$~K\@. We show that GdPd$_3$ undergoes two distinct magnetic transitions at $T_{\rm N1} = 6.5$~K and $T_{\rm N2} = 5.0$~K, respectively. The $\chi(T)$  and magnetization $M$ versus $H$ isotherm data along with the MR data show that the spin structure of the Gd spins below $T_{\rm N2}$ is fragile and can be significantly altered by relatively small $H$. The fragile spin structure of the compound results in a cascade of field-induced spin-reorientation transitions. Our results show that the oscillating MR below $T_{\rm N2}$ reflects each field-induced spin-reorientation transition that the system undergoes in a varying $H$.\\

\section*{Results}
\subsection*{Magnetoresistance.}
\begin{figure}[h]
\centering
\includegraphics[width=6.2in]{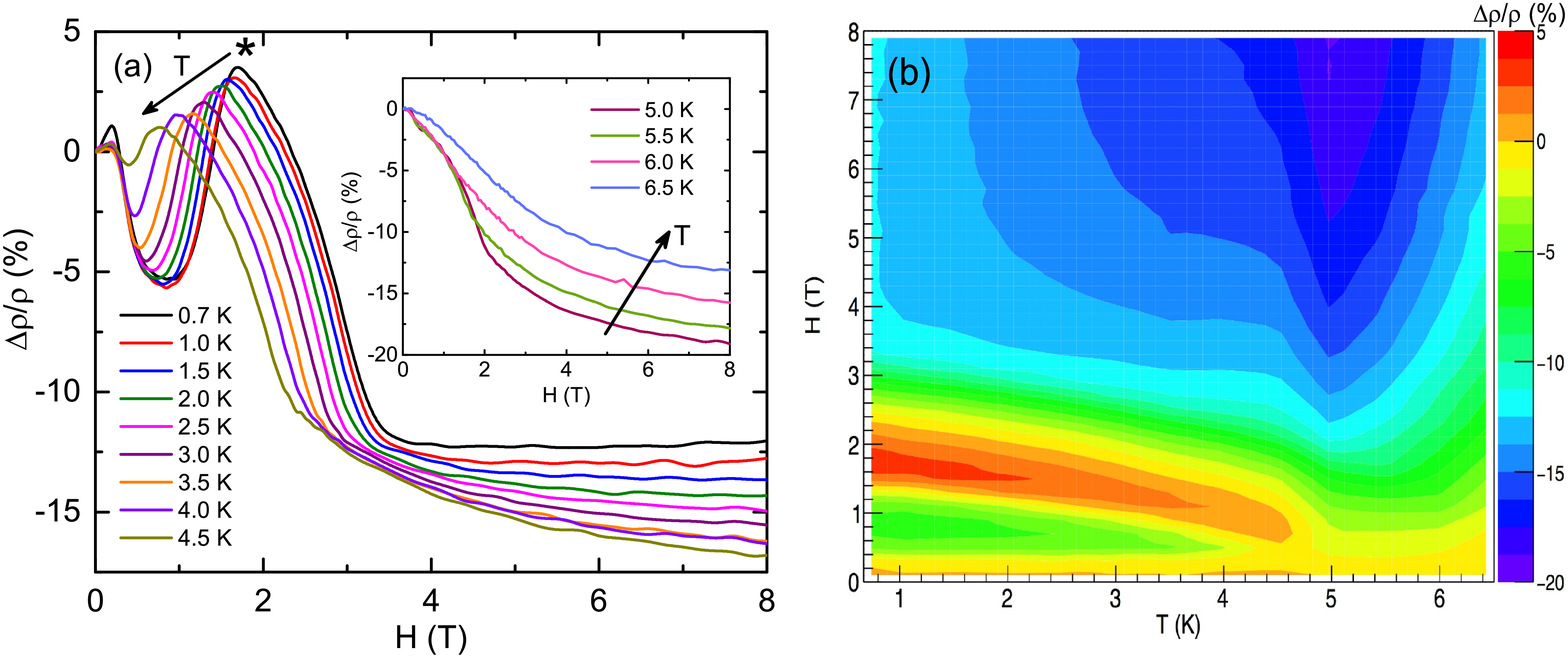}
\caption{(a) Magnetoresistance  $\Delta\rho/\rho$ versus applied magnetic field $H$ for GdPd$_3$ measured at nine different temperatures $T$ between 0.7 and 4.5~K\@. The peak with the highest positive MR is indicated with an asterisk. Inset:  $\Delta\rho/\rho$ versus $H$ at four different $T$'s  between 5 and 6.5~K\@. The arrows in the figure as well as in the inset indicate increasing temperatures of the isotherms. (b) The $\Delta\rho/\rho$ of GdPd$_3$ depicted in a $H$-$T$ color contour plot.}
\label{fig:Figure_MR-1}
\end{figure}

The field dependences of the low-temperature MR $\equiv \Delta\rho/\rho=[\rho(H)-\rho(0)]/\rho(0)$ measured at thirteen different temperatures between 0.7 and 6.5~K are shown in Fig.~\ref{fig:Figure_MR-1}(a) and the inset therein. While the data below $T_{\rm N2}$ show oscillating behavior, the data for $T \ge T_{\rm N2}$ exhibit a negative MR which monotonically decreases with the increase of $H$ up to the maximum $H = 8$~T of the measurement. The novel oscillating behavior of MR is depicted in a $H-T$ color contour plot [Fig.~\ref{fig:Figure_MR-1}(b)], highlighting the regions of nearly the same values and the crossovers between the positive and negative MR's. 

\begin{figure}[H]
\centering
\includegraphics[width=6.2in]{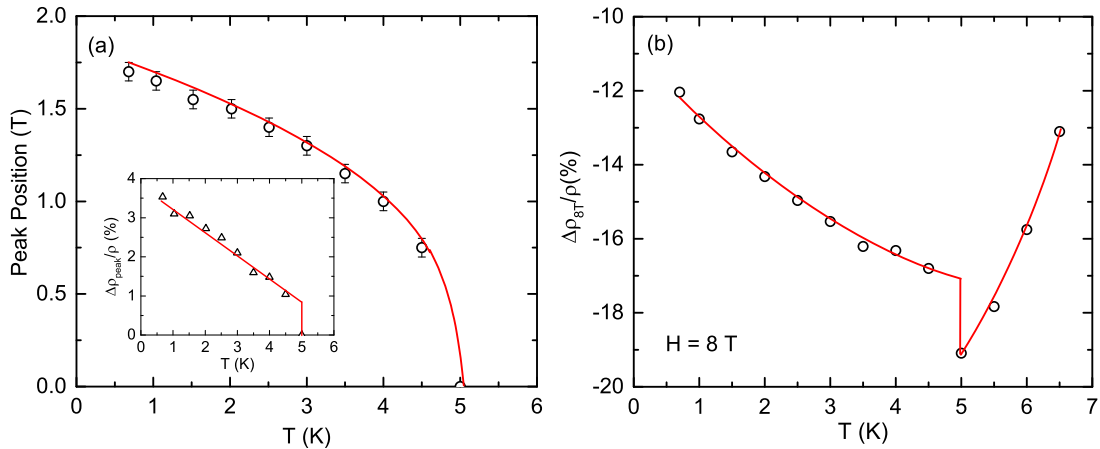}
\caption{(a) Temperature $T$ dependence of the position of the largest positive magnetoresistance (MR) peak of GdPd$_3$ marked with an asterisk in Fig.~\ref{fig:Figure_MR-1}(a). Inset: $T$ dependence of the value of the positive MR peak marked with the asterisk. (b)  $T$ dependence of MR of GdPd$_3$ at applied filed $H=8$~T\@. Solid curves/lines in both figures as well as in the inset are guides to the  eye.}
\label{fig:Figure_MR-2}
\end{figure}

The general features of the MR data for $T < T_{\rm N2}$ are quite similar, thus we use the lowest $T$ data at 0.7~K in this $T$ range to discuss their characteristics in the following. The MR shows a small positive peak centered at 0.2~T\@. The increase of $H$ turns MR negative and results in a local minimum whose position and depth depends on the temperature. At 0.7~K the minimum occurs at $\sim 1$~T. The further increase of $H$ results in a positive MR at 1.4~T and a second maximum located at 1.7~T\@. Increasing the $H$ even further results in a nearly monotonic decrease of MR that turns negative at 2.3~T and shows a plateau or tendency to saturation above $\sim 3.5$~T\@. The variation of the position of the positive MR peak [marked with an asterisk in Fig.~\ref{fig:Figure_MR-1}(a)] with $T$ is shown in Fig.~\ref{fig:Figure_MR-2}(a). The data show that with the increase of $T$ the peak position monotonically shifts to lower $H$ values and the peak finally disappears at 5~K. The peak MR exhibits a nearly linear decrease with the increase of $T$ before attaining a zero value at 5~K [Inset, Fig.~\ref{fig:Figure_MR-2}(a)].  The $T$ dependence of the MR at 8~T ($\Delta\rho_{\rm 8T}/\rho$) exhibits a monotonic decrease in the value before undergoing a discontinuous jump at 5~K, after which the data again show a monotonic behavior but this time the MR increases with the increase of $T$ [Fig.~\ref{fig:Figure_MR-2}(b)]. We return to the analysis and interpretation of the MR data of GdPd$_3$ in the discussion section. \\

\subsection*{Magnetic susceptibility and magnetization versus field isotherms.}

\begin{figure}[ht]
\centering
\includegraphics[width=6.2in]{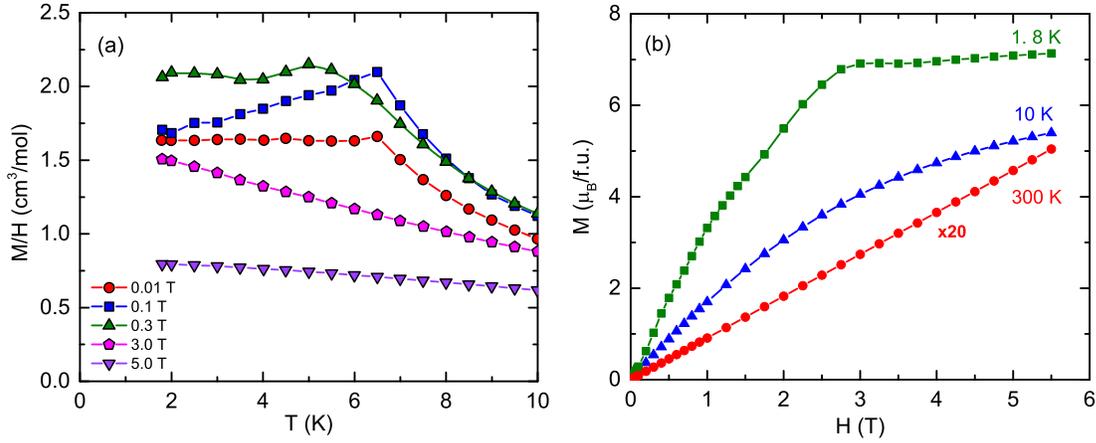}
\caption{(a) Zero-field-cooled magnetic susceptibility $\chi \equiv M/H$ of GdPd$_3$ versus temperature $T$ measured in five different applied magnetic fields $H$ between 0.01 and 5~T\@. (b) Variation of the the isothermal magnetization $M$ with $H$ measured at $T = 1.8$, 10 and 300 K\@. For better visibility, the data at 300~K are multiplied by 20. The solid curves in both figures are guides to eye.}
\label{fig:Figure_Mag-1}
\end{figure} 

Low-temperature $\chi(T) \equiv M/H$ data of GdPd$_3$ at five different $H$'s between 0.01 and 5~T are shown in Fig.~\ref{fig:Figure_Mag-1}(a). It is evident from the figure that the value of $\chi$ and the nature of its $T$ dependence depend sensitively on $H$. The value of $T_{\rm N}$ along with the parameters $\xi$ and $ f = \theta_{\rm p}/T_{\rm N}$ calculated at different $H$'s are listed in Table~\ref{Table}. The $\chi(T)$ measured at 0.01~T shows a kink at $T_{\rm N1} = 6.5$~K, below which it is nearly $T$ independent. This kind of $\chi(T)$ behavior below $T_{\rm N}$ is expected for frustrated 120$^\circ$-triangular lattice antiferromagnets\cite{Johnston-2012,Brown-2006, Kadowaki-1995, Hirakawa-1983}. However, the data at 0.1~T show strikingly different characteristics where the $\chi(T)$ shows a kink at the same $T_{\rm N1} = 6.5$~K, but below this temperature the $\chi$ monotonically decreases with the decrease of $T$. The observed $T$ dependence of $\chi$ below $T_{\rm N1}$ and the value of $\xi = 0.81$ at 0.1~T suggest a noncollinear AFM spin structure in the compound \cite{Johnston-2012,Johnston-2015,Cho-2005}. The $\chi(T)$ measured at higher $H = 0.3$~T again shows nearly $T$-independent behavior with $\xi = 0.96$ below a ordering temperature which is reduced to a value of $T_{\rm N}(0.3~{\rm T}) = 5$~K (Table~\ref{Table}). The ordering temperature of AFM's usually decreases with increasing $H$. However, in the case of GdPd$_3$ the value of $T_{\rm N}$(0.3~T) coincides with the spin-reorientation transition temperature $T_{\rm N2}$ indicated from the $C_{\rm p}(T)$ and $\rho(T)$ data discussed below. At even higher fields, the transition in the $\chi(T)$ data completely disappears [Fig.~\ref{fig:Figure_Mag-1}(a)]. The following conclusions can be drawn from the $\chi(T)$ data of GdPd$_3$; the spin structure of the compound is (i) noncollinear and (ii) highly fragile. The latter inference is established from the remarkable change in the $T$ dependence of $\chi$ between relatively low applied fields of 0.01 and 0.1~T. 

\begin{table}[ht]
\centering
	\begin{tabular}{|l| c| c| c|}
		\hline
		 $H (\rm T)$ & $T_{\rm N}(H)$ & $\xi = \chi(0)/\chi(T_{\rm N})$ &  $f = \theta_{\rm p}/T_{\rm N}$\\
	 \hline
	 		 0.01 & 6.5~K [$T_{\rm N1}$] & 0.99(1) &  0.9(3) \\
			 0.1 & 6.5~K [$T_{\rm N1}$] & 0.81(1) & 0.9(3)\\
			 0.3 & 5.0~K [$T_{\rm N}$(0.3 T)] & 0.96(1) & 1.2(4) \\
			\hline
		\end{tabular}
\caption{\label{Table} Magnetic ordering temperature $T_{\rm N}$ deduced from the $\chi(T)$ measurements, $\xi = \chi(0)/\chi(T_{\rm N})$ and  $f = \theta_{\rm p}/T_{\rm N}$ calculated at three different applied fields $H$. Whenever there is an obvious peak (or kink) in the $\chi(T)$ data, the $T_{\rm N}$ is taken as the peak (or kink) temperature. At higher $H$'s where there is no obvious kink, the  $T_{\rm N}$ is taken as the $T$ where the change in slope of $\chi(T)$ is maximum. The $\chi$ at 1.8~{\rm K} is taken as $\chi(0)$. The value of the Weiss temperature $\theta_{\rm p}$ in the Curie-Weiss law for GdPd$_3$ at $T > T_{\rm N1}$ is $+6(2)$~K\@.}
\end{table} 

The isothermal magnetization $M$ versus $H$ data taken at 1.8, 10 and 300~K are shown in Fig~\ref{fig:Figure_Mag-1}(b). The $M(H)$ data at 1.8~K show a monotonic but nonlinear increase of $M$ with $H$ below 3~T. The data indicate multiple field-induced spin-reorientation transitions that are evident from the change of the slope of the $M$ versus $H$ plot at 1.8~K. We return to this point and elaborate in the discussion section. The data at 1.8~K exhibit saturation at $\sim 3$~T to a value $\mu_{\rm sat} = gS\mu_{\rm B} = 7$~$\mu_{\rm B}$ expected for a $S = 7/2$ Gd$^{+3}$ ions considering the spectroscopic splitting factor $g =2$. The $M(H)$ plot at 10~K shows a monotonic and nonlinear increase of $M$ with $H$ as expected in the paramagnetic (PM) state at $T > T_{\rm N}$. The $M(H)$ data at 300~K show a linear behavior as expected for a compound in the PM state at $T \gg \mu_{\rm sat}H/k_{\rm B}$, where $k_{\rm B}$ is Boltzmann's constant. \\

\subsection*{Heat capacity.}

\begin{figure}[ht]
\centering
\includegraphics[width=6.2in]{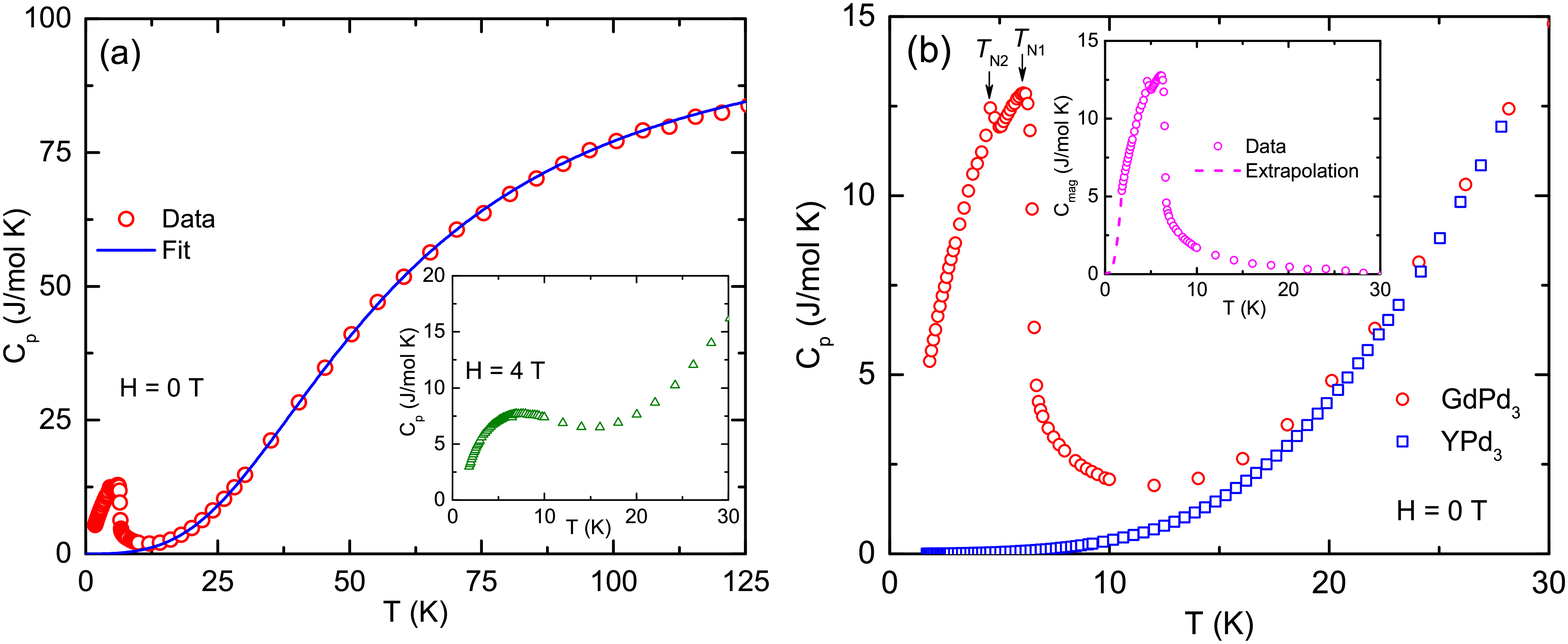}
\caption{(a) Molar heat capacity $C_{\rm p}$ of GdPd$_3$ versus temperature $T$. The solid blue curve is a fit using by Eq.~\ref{Eq:HC}. Inset: $C_{\rm p}(T)$ measured at $H = 4$~T\@. (b) $C_{\rm p}(T)$ for GdPd$_3$ and its nonmagnetic analog YPd$_3$ at low temperatures. The $C_{\rm p}(T^{*})$ data of YPd$_3$ incorporates the effect of the molar mass difference of the two compounds. Inset: $T$ dependence of magnetic part of the heat capacity, $C_{\rm mag} = C_{\rm{p\,GdPd}_3}-C_{\rm{p\,YPd}_3}$. The dashed line in the inset for $T \le 1.8$~K is an extrapolation $C_{\rm extrap}= BT^3$.} 
\label{fig:Figure_HC-1}
\end{figure}

The $C_{\rm p}(T)$ data for GdPd$_3$ taken at $H = 0$ are shown in Fig.~\ref{fig:Figure_HC-1}(a). The data show an upturn below $\sim 10$~K and exhibit two humps centered $T = 5.0$ and 6.5~K [Fig.~\ref{fig:Figure_HC-1}(b)], respectively. While the anomaly at 6.5~K reflects the $T_{\rm N1}$ of the $\chi(T)$ data measured at 0.01~T, the feature at $T_{\rm N2} = 5.0$~K is most likely due to a zero field spin-reorientation transition which incidentally coincides with the $T_{\rm N}$(0.3~T)  in $\chi(T)$. It is interesting that while the $C_{\rm p}(T)$ data clearly capture two magnetic transitions, the $\chi(T)$ data at lower fields (0.01 and 0.1~T) do not show any signature of the lower-$T$ transition at $T_{\rm N2}$. An applied $H$ of 4~T masks the two distinct transitions observed in $H = 0$ and instead results in a broad hump in $C_{\rm p}(T)$ [inset, Fig.~\ref{fig:Figure_HC-1}(a)]. This observation is consistent with the $\chi(T)$ data taken at $H = 3$~T and 5~T that show no evidence for a transition  [Fig.~\ref{fig:Figure_Mag-1}(a)]. 

We fitted the $C_{\rm p}(T)$ data above 20~K by
\begin{equation}
C_{\rm p}(T) = \gamma T + nC_{\rm{V\,Debye}}(T) ,
\label{Eq:HC}
\end{equation}
where $\gamma$ is the Sommerfeld coefficient, $n$ is the number of atoms per formula unit which is 4 for GdPd$_3$ and $C_{\rm{V\,Debye}}$ is the Debye molar lattice heat capacity at constant volume \cite{Kittel-2005} described by
\begin{equation}
C_{\rm{V\,Debye}}(T) = 9 R \left( \frac{T}{\Theta_{\rm{D}}} \right)^3 {\int_0^{\Theta_{\rm{D}}/T} \frac{x^4 e^x}{(e^x-1)^2}\,dx},
\label{Eq:DebyeHC}
\end{equation}

\begin{figure}[ht]
\centering
\includegraphics[width=3.5in]{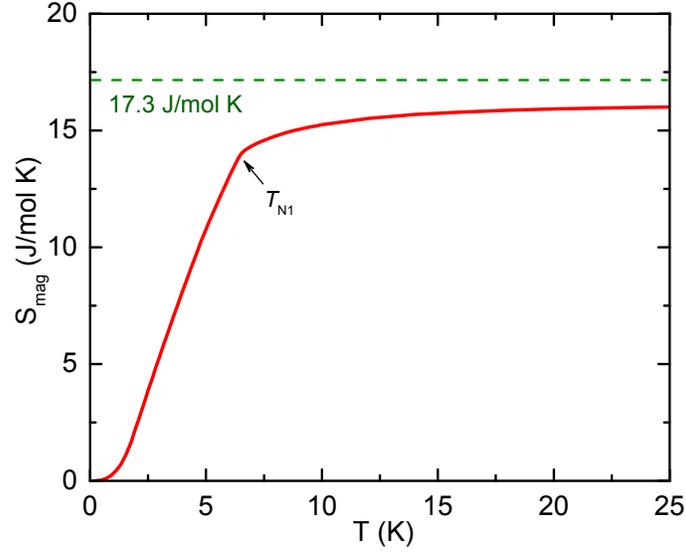}
\caption{Magnetic entropy $S_{\rm mag}$ of GdPd$_3$ versus temperature $T$\@. The horizontal dashed green line shows the value of $S_{\rm mag}$ expected for spins $S = 7/2$, $S_{\rm mag} (T \rightarrow \infty)= R{\rm ln}8$.}
\label{fig:Figure_HC-2}
\end{figure}

where $\Theta_{\rm D}$ is Debye temperature and $R$ is the molar gas constant. The data were fitted using Eqs.~(\ref{Eq:HC}) and (\ref{Eq:DebyeHC}) employing the Pad\'e fitting function described in ref.~\citenum{Goetsch-2012}. A good fit to the data for $20\le T \le 125$~K was obtained with the fitted values of the parameters $\gamma = 7(1)$~mJ/mol K and $\Theta_{\rm D} = 237(1)$~K [Fig.~\ref{fig:Figure_HC-1}(a)]. 

To estimate the magnetic contribution to $C_{\rm p}(T)$ of GdPd$_3$ we used the $C_{\rm p}(T)$ of YPd$_3$ as the nonmagnetic reference data for the former. YPd$_3$ has the same crystal structure as GdPd$_3$ and has nearly the same lattice parameter $a = 4.069$~\AA~(ref.~\citenum{Pandey-2009b}), but the molar masses of the two compounds differ by about 14\%. The $\Theta_{\rm D}$ depends on the molar mass $M_{\rm mol}$ of a system as $\sim 1/M_{\rm mol}^{1/2}$ and the Debye lattice heat capacity is a function of $T/\Theta_{\rm D}$. Thus to compensate the effect of the molar mass difference between the two compounds, the $T$-axis of YPd$_3$ was scaled using the following expression,
\begin{equation}
T^{*} =  \frac{T}{[{M_{\rm GdPd_3}/M_{\rm YPd_3}}]^{1/2}},
\label{Eq:MolarMass}
\end{equation}
The low-temperature $C_{\rm p}(T)$ of GdPd$_3$ is replotted in Fig.~\ref{fig:Figure_HC-1}(b) along with the $C_{\rm p}(T^{*})$ data of YPd$_3$. The magnetic contribution $C_{\rm mag}$ to the $C_{\rm p}$ of GdPd$_3$, $C_{\rm mag} = C_{\rm p\,GdPd_3} -  C_{\rm p\,YPd_3}$, is plotted versus $T$ in the inset of Fig.~\ref{fig:Figure_HC-1}(b). The $C_{\rm mag}(T)$ is sizable at 10~K, which is consistent with the $M(H)$ data taken at the same temperature [Fig.~\ref{fig:Figure_Mag-1}(b)], and becomes negligibly small above $\sim 20$~K. These features correlate very well with the $\rho(T)$ data discussed below. The magnetic contribution $S_{\rm mag}$ to the entropy of a system can be estimated from the $C_{\rm mag} $ data using the expression
\begin{equation}
S_{\rm mag}(T) = \int_{0}^{T}{\frac{C_{\rm mag}(T)}{T}}dT.
\end{equation}  
The calculated $S_{\rm mag}$ versus $T$ is plotted in Fig.~\ref{fig:Figure_HC-2}. The high-$T$ limit expected for $S = 7/2$ Gd$^{+3}$ ions, $S_{\rm mag} (T \rightarrow \infty) = R{\rm ln}(2S+1) = R{\rm ln}8$ = 17.3~J/mol~K, is indicated in the figure. The  $S_{\rm mag}(T)$ undergoes a sharp change at $T_{\rm N1} = 6.5$~K and above this temperature shows a tendency for saturation to the limiting value which is nearly attained at $\sim 20$~K. The somewhat smaller value of $S_{\rm mag}$ than the expected high-$T$ limit is likely due to inaccuracy in the lattice contribution to $C_{\rm p}(T)$. The entropy change above $T_{\rm N1}$ arises from short-range dynamic AFM ordering of the Gd spins. \\

\subsection*{Electrical resistivity.}

\begin{figure}[ht]
\centering
\includegraphics[width=6.2in]{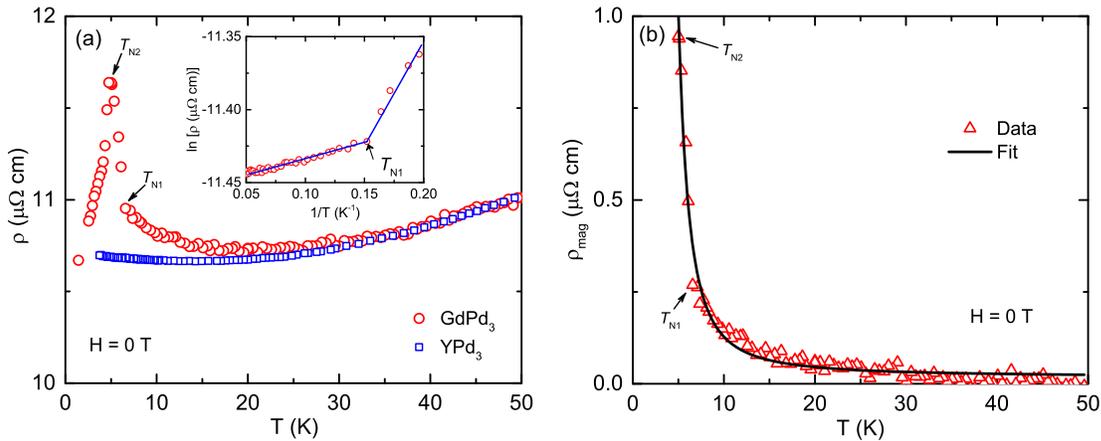}
\caption{(a) Electrical resistivity $\rho$ of GdPd$_3$ versus temperature $T$ plotted along with the data for YPd$_3$. The $\rho(T)$ of the latter has been scaled by multiplying by a constant so that the data at higher-$T$'s overlap with those of the former. Inset: The $\rho(T)$  data above the peak at $T_{\rm N2}$, between 5 and 20~K, plotted as ln$\rho$ versus $1/T$. The solid blue lines are guides to the eye. (b) Magnetic contribution  to the resistivity of GdPd$_3$ $\rho_{\rm mag} = \rho_{\rm GdPd_3} - \rho_{\rm YPd_3}$ versus $T$  for $T \ge 5$~K\@. The solid black curve is the fit of the data by Eq.~\ref{Eq:rhoFit}. The transition temperatures $T_{\rm N1}$ and $T_{\rm N2}$ are indicated by black arrows in both figures.}
\label{fig:Figure_Res}
\end{figure}

The $\rho(T)$ of GdPd$_3$ for $T \le 50$~K is plotted along with the data for the nonmagnetic analogue YPd$_3$ in Fig.~\ref{fig:Figure_Res}(a). The $\rho(T)$ data between 0.6 to 300~K at $H = 0$~T and 0.7 to 150~K at $H = 8$~T are plotted in Fig.~2 of the supplemental material. Similar to the $C_{\rm p}(T)$ data discussed above, the $\rho(T)$ of YPd$_3$ qualitatively describes the behavior of GdPd$_3$ for $ T\gtrsim 20$~K\@. The $\rho(T)$ of GdPd$_3$ exhibits a sharp increase with the increase of $T$ and exhibits a narrow peak at $T_{\rm N2}$, above which it sharply decreases with increasing $T$ and undergoes a change in slope at $T_{\rm N1}$. To highlight the latter we plotted ln$\rho(T)$ versus $T^{-1}$ in the inset, which clearly shows a change in slope at $6.5$~K\@. The upturn below $\sim 20$~K in the $\rho(T)$ is likely due to the opening of an AFM superzone pseudogap at the Fermi surface due to emergence of an incommensurate AFM ordering and a superzone gap at $T_{\rm N1}$ (refs.~\citenum{Jensen-1991,Elliott-1964,Ellerby-1998,Park-2005, Wilding-1965, Mackintosh-1962}). The sharp decrease in $\rho(T)$ below $T_{\rm N2}$ is evidently due to a steep decrease in the spin-disorder scattering below this temperature. 

To further explore this scenario we fitted the overall $T$-dependence of the magnetic contribution to the resistivity $\rho_{\rm mag}$ of GdPd$_3$ for $T \ge 5$~K by the activated behavior 
\begin{equation}
\rho(T) =  Ae^{-\Delta/k_{\rm B}T},
\label{Eq:rhoFit}
\end{equation}
where $2\Delta$ is the superzone band gap and $A$ is a constant. We obtained a reasonably good fit to the data for $T \ge 5$~K with $\Delta = 20.7(4)$~K and $A = 0.016(1)$~$\mu\Omega$-cm [Fig.~\ref{fig:Figure_Res}(b)]. The quality of the fit is quite good between $T_{\rm N2}$ and $T_{\rm N1}$, but it decreases between $T_{\rm N1}$ and  20~K. However the effect of the kink at $T_{\rm N1}$ is small compared to the activated increase observed in $\rho_{\rm mag}$, thus the data can still be reasonably fitted using a single parameter $\Delta$. The $\rho_{\rm mag}(T)$ data presented here clearly show the existence of a superzone pseudogap for $T \ge T_{\rm N1}$ and a gap for $T < T_{\rm N1}$ at the Fermi surface. Because the gap and pseudogap are associated with the conduction electrons with a heat capacity of order $\le 0.01$~J/mol~K below 20~K (see Figure~3 of supplemental material) the changes in $C_{\rm p}$ due to the opening of the gap and pseudogap are too small to resolve in the Fig.~\ref{fig:Figure_HC-1}(b) because the $C_{\rm p}$ is strongly dominated by the magnetic contribution.\\

\section*{Discussion}

\begin{figure}[ht]
\centering
\includegraphics[width=3.5in]{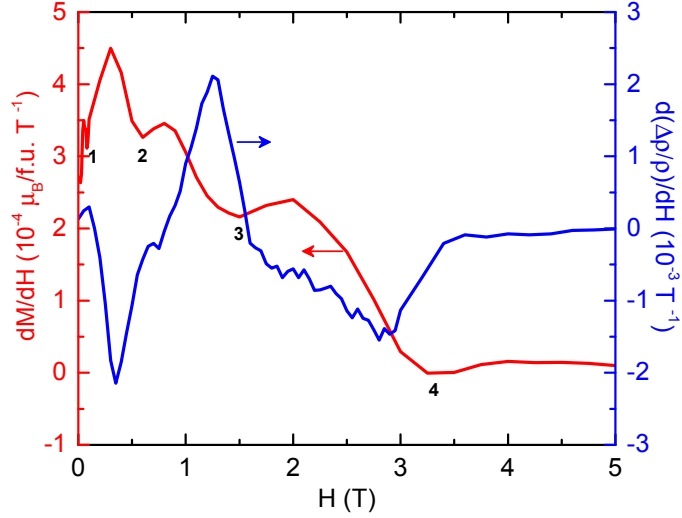}
\caption{Magnetic field $H$ derivative of the isothermal magnetization d$M$/d$H$ versus applied magnetic field $H$ of GdPd$_3$ at 1.8 K (left ordinate) and the $H$ derivative of magnetoresistance ${\rm d}(\Delta\rho/\rho)/{\rm d}H$ versus $H$ of GdPd$_3$ at 1.5 K (right ordinate). Four distinct minima observed in the d$M$/d$H$ versus $H$ plot are indicated by numbers 1, 2, 3 and 4, respectively, in the order their occurrence with increasing $H$.}
\label{fig:Figure_Derivative}
\end{figure}

The positive value of $\theta_{\rm p}$ of GdPd$_3$ (Table~\ref{Table}) suggests sizable presence of ferromagnetic (FM) interactions in the material. On the other hand, the nature of $\chi$ versus $T$ plot at low fields and the value of $\xi = \chi(0)/\chi(T_{\rm N}) $, which is not close to 2/3 expected for a polycrystalline sample of a collinear AFM, indicate that the magnetic spin structure is noncollinear\cite{Sanchez-1976} and fragile, which can be significantly altered by relatively small $H$. It is interesting that while the $C_{\rm p}(T)$ and $\rho(T)$ data clearly show two magnetic transitions at $T_{\rm N1}$ and $T_{\rm N2}$, the $\chi(T)$ data at small $H$ show only one transition at $T_{\rm N1}$. The $C_{\rm p}(T)$, $\rho(T)$ and $M(H)$ data together show that significant short-range magnetic correlations persist in the system above $T_{\rm N1}$ up to $\sim 20$~K. The low-$T$ $\rho(T)$ data at $H = 0$ clearly indicate the opening of a superzone gap (pseudogap for $T \ge T_{\rm N1}$ and a gap for $T < T_{\rm N1}$) at the Fermi surface, which is a manifestation of a magnetic structure whose periodicity is incommensurate with the periodicity of crystal lattice. The $\rho_{\rm mag}$ versus $T$ data show that the effect of opening of the superzone gap in this system can be modeled using a simple thermally-activated single band gap expression. The features in $C_{\rm mag}$ and $\rho$ at $T_{\rm N2}$ apparently arise from a spin reorientation transition. The $\rho_{\rm GdPd_3}$ approaches the $\rho_{\rm YPd_3}$ for $T < T_{\rm N2}$, indicating that the decrease in $\rho_{\rm GdPd_3}$ below $T_{\rm N2}$ is due to the the loss of spin-disorder scattering below this temperature. 

To clarify the driving mechanism for the observed novel oscillating MR behavior between positive and negative values we have plotted in Fig.~\ref{fig:Figure_Derivative} the $H$ dependence of the derivative of the isothermal magnetization $M' = {\rm d}M/{\rm d}H$ at 1.8~K from Fig.~\ref{fig:Figure_Mag-1}(b) along with the derivative of the MR data  ${\rm d}(\Delta\rho/\rho)/{\rm d}H$ taken at 1.5~K from Fig.~\ref{fig:Figure_MR-1}(a). The $M'$ versus $H$ plot shows a cascade of steep decreases followed by shallow minima with increasing $H$. The four shallow minima observed in the measured $H$ range are marked in Fig.~\ref{fig:Figure_Derivative} by the numbers 1, 2, 3 and 4, respectively. This observation clearly shows that GdPd$_3$ undergoes several $H$-dependent spin reorientation transitions--a behavior which is apparently a manifestation of the presence of competing AFM interactions and a significant FM interaction in the system. 

The two main conclusions that can be drawn from the plots shown in Fig.~\ref{fig:Figure_Derivative} are; (i) the difference in $H$ between two successive minima as well as the length of the plateau that appears following the minima in the $M'(H)$ plot increase with increasing $H$ and (ii) the $M'$ and ${\rm d}(\Delta\rho/\rho)/{\rm d}H$ are strongly inversely correlated to each other, {\it i.e.}, when the former increases the later decreases and when the former exhibits a peak the latter shows a dip. Figure~\ref{fig:Figure_Derivative} shows that even a small feature in the $M(H)$ data, for example the minimum marked by ``2", leaves it's signature in the $\Delta\rho/\rho$ data. Such a strong correlation between two properties measured in two entirely different measurements, where the former [$M(H)$] is a thermodynamic measurement and the latter [$\Delta\rho/\rho$] is a transport one, is certainly a rare occurrence. The FM correlations lead to a negative MR while the AFM correlations usually result in a positive MR \cite{Yamada-1973}. An increase in $M'$ with increasing $H$ indicates the field-induced growth of the FM component in the system and manifests in the decrease of ${\rm d}(\Delta\rho/\rho)/{\rm d}H$, while a decrease in $M'$ with increasing $H$ or a plateau suggests a halt in the growth of the FM component and thus results in an increase of ${\rm d}(\Delta\rho/\rho)/{\rm d}H$.  We propose that the competing AFM and FM interactions and the resultant extremely field-sensitive fragile spin structure of GdPd$_3$ cause the observed novel oscillating behavior of the MR. 

Metallic GdPd$_3$ is the simplest system (binary system), crystallizing in the simplest structure (primitive cubic structure) and the magnetism of the compound is due to the simplest rare-earth ion ($S$-state Gd$^{+3}$-ion). However, the compound exhibits complicated magnetic, electrical and magnetotransport phenomena. These evidences of fragile magnetism indicate that it would be very interesting to experimentally investigate the evolution the spin structure of GdPd$_3$ in the presence of $H$. Due to the low values of $T_{\rm N1}$ and $T_{\rm N2}$, it is plausible that the magnetic dipole interactions\cite{Johnston-2016} may compete with RKKY interactions to determine the magnetic structure of the compound. During the review of this manuscript, we became aware of two recent works\cite{Zhang-2015, Huang-2016} that report sample- and relative orientation between magnetic field and current-dependent chirality-driven  oscillating magnetoresistance between positive and negative values in TaAs. The underlying mechanisms of the oscillating MR in TaAs and GdPd$_3$ are however very different. While the origin of the observed negative MR in the nonmagnetic Weyl semimetal TaAs has been attributed to the chirality anomaly, the oscillating MR in the magnetic metal GdPd$_3$ is shown to be driven by the underlying fragile spin structure of the material. 

In conclusion, we have investigated the novel oscillating MR observed in metallic GdPd$_3$ below the magnetic ordering temperature. The $\chi(T)$ at low fields ($H \le 0.1$~T) shows a magnetic transition at $T_{\rm N1} = 6.5$~K. The value of $\xi = \chi(0)/\chi(T_{\rm N})$ is $H$-dependent and is significantly higher than 2/3 expected for a polycrystalline collinear AFM, suggesting a noncollinear spin arrangement in the material. It is indeed interesting that while the $\chi(T)$ shows only one magnetic transition at a particular $H$, the $C_{\rm p}(T)$ and $\rho(T)$ data at $H = 0$ clearly show the presence of two distinct transitions at $T_{\rm N1} = 6.5$~K and $T_{\rm N2} = 5.0$~K\@. The $\rho(T)$ data show the existence of a magnetic superzone gap below $T_{\rm N1}$ that arises from a magnetic structure incommensurate with the periodicity of the crystal lattice. This observation suggests that the underlying spin structure of GdPd$_3$ is noncollinear as well as incommensurate to the periodicity of the crystal lattice below $T_{\rm N1}$. The $\chi(T)$ and $M(H)$ data along with the MR data suggest that the spin structure of the compound below $T_{\rm N2}$ is fragile and can be significantly modified by a small $H$. The $M(H)$ isotherm at 1.8~K suggests the presence of several $H$-induced spin reorientation transitions. The features observed in the oscillating MR correlate very well with the positions and the nature of the spin reorientation transitions, thus evidently are a manifestation of them. The observed delicate correlation between the two properties--magnetization and magnetoresistance, where the former is a thermodynamic property while the latter is a transport one, is a rare occurrence. The rich magnetotransport characteristics of GdPd$_3$ have prospects for applications in field-sensitive devices. Such applications become more plausible if the strength of the MR oscillations and temperature below which the oscillations are observe could be increased using single-crystal variants or by perturbations such as chemical substitution. Studies on single-crystal samples of GdPd$_3$ might be helpful to determine if domain-wall motion and/or domain reorientation effects are relevant to our MR results. Additionally, the probable reduction of impurity scattering and grain-boundary effects in the single-crystal samples may lead to  enhancement of the observed oscillations. It would also be exciting to investigate the MR characteristics of GdPd$_3$ in disordered and/or epitaxial thin film forms. The change in dimensionality usually has a significant effect on the electrical transport properties. The promising MR properties of GdPd$_3$ encountered in the bulk form stimulate such studies that might lead to exciting outcomes.\\

\section*{Methods}
A polycrystalline sample of GdPd$_3$ was synthesized by arc-melting the stoichiometric amount of highly pure ($\ge 99.9\%$) constituent elements under argon followed by vacuum annealing for 240~h at 1000~$^{\circ}$C (ref.~\citenum{Pandey-2009a}). Powder x-ray diffraction data taken at room temperature (see Figure-1 of supplemental material) and their Rietveld refinement \cite{Carvajal-1993} suggest that the synthesized compound is a single phase and is free from any detectable impurity \cite{Pandey-2008}. The refined value of the cubic lattice parameter $a$ is 4.0919(4)~\AA.\@. Temperature- and magnetic field-dependent electrical transport measurements were carried out using the four-probe technique in a Quantum Design Physical Properties Measurement System (PPMS) equipped with a $^3$He refrigeration system. The MR data do not show any significant dependence on the relative orientation between the current direction and $H$. Heat capacity $C_{\rm p}(T)$ was measured by relaxation measurement in the PPMS. The temperature dependence of $\chi$  and field dependence of the magnetization $M$ was measured in a Quantum Design Magnetic Properties Measurement System (MPMS). The $\chi(T)$ data were taken in both zero-field-cooled (ZFC) and field-cooled (FC) conditions at the lowest field $H = 0.01$~T\@. Since the ZFC and FC data overlap with each other in the entire $T$ range of the measurement at this $H$, the data at higher $H$'s were taken only in the ZFC condition. The overlapping ZFC and FC data suggest that our sample is free from blocking and pinning effects.\\

\section*{Acknowledgments}
A.P. acknowledges T. Thakore, A. Saleheen and N. S. Sangeetha for assistance. The authors thank P. W. Adams of LSU and I. Das of SINP for helpful discussions and suggestions. The materials were prepared at SINP. The work at SINP was funded by Department of Atomic Energy, India. The magnetic and heat capacity measurements were carried out at the Ames Laboratory. The work at Ames Laboratory was supported by the U.S.~Department of Energy, Office of Basic Energy Sciences, Division of Materials Sciences and Engineering.  Ames Laboratory is operated for the U.S.~Department of Energy by Iowa State University under Contract No.~DE-AC02-07CH11358. The temperature and field dependent electrical and magnetotransport measurements were carried out at LSU. The work at LSU was supported by Department of Energy, Office of Science, Basic Energy Sciences under Award No.~DE-FG02-07ER46420.\\

\section*{Author Contributions}
A.P. and C.M. synthesized the sample. A.P. carried out the measurements, coordinated the project and wrote the first daft of the manuscript. All authors contributed to data analysis, interpretation and to the final version of the manuscript. \\

\section*{Additional Information}
The authors declare no competing financial interests. \\

\noindent AP's Present address: Department of Physics and Astronomy, Texas A\& M University, College Station, Texas 77840-4242, USA.


\begin{thebibliography}{99}

\bibitem{Moritomo-1996} Moritomo, Y. {\it et al.} Giant magnetoresistance of manganese oxides with a layered perovskite structure. {\it Nature} {\bf 380}, 141-144 (1996).

\bibitem{Husmann-2002} Husmann, A.  {\it et al.}  Megagauss sensors. {\it Nature} {\bf 417}, 421-424 (2002).

\bibitem{Lenz-1990} Lenz, J. E. A review of magnetic sensors. {\it Proc. IEEE}  {\bf 78}, 973-989 (1990).

\bibitem{Salamon-2001} Salamon, M. B. \& Jaime, M. The physics of manganites: Structure and transport. {\it Rev. Mod. Phys.} {\bf 73}, 583-628 (1997).

\bibitem{Ramirez-1997} Ramirez, A. P. Colossal magnetoresistance. {\it J. Phys.: Condens. Matter} {\bf 9}, 8171-8199 (1997).

\bibitem{Baibich-1988} Baibich, M. N. {\it et al.} Giant magnetoresistance of (001)Fe/(001)Cr magnetic superlattices. {\it Phys. Rev. Lett.} {\bf 61}, 2472-2475 (1988).

\bibitem{Binasch-1989} Binasch, G. {\it et al.} Enhanced magnetoresistance in layered magnetic structures with antiferromagnetic interlayer exchange. {\it Phys. Rev. B} {\bf 39}, 4828(R)-4830(R) (1989).

\bibitem{Ali-2014} Ali, M. N. {\it et al.} Large, non-saturating magnetoresistance in WTe$_2$. {\it Nature} {\bf 514}, 205-208 (2014).

\bibitem{Tafti-2016} Tafti, F. F. {\it et al.} Resistivity plateau and extreme magnetoresistance in LaSb. {\it Nature Phys.} {\bf 12}, 272-277 (2016).

\bibitem{Liang-2015} Liang, T. {\it et al.} Ultrahigh mobility and giant magnetoresistance in the Dirac semimetal Cd$_3$As$_2$. {\it Nature Mater.} {\bf 14}, 280-284 (2015).

\bibitem{Wang-2014} Wang, K. {\it et al.} Anisotropic giant magnetoresistance in NbSb$_2$. {\it Sci. Rep.} {\bf 4}, 7328:1-6 (2014).

\bibitem{Wang-2015} Wang, L. {\it et al.} Tuning magnetotransport in a compensated semimetal at the atomic scale. {\it Nat. Commun.} {\bf 6}, 8892: 1-7 (2015).

\bibitem{Keum-2015} Keum, D. H. {\it et al.} Bandgap opening in few-layered monoclinic MoTe$_2$. {\it Nat. Phys.} {\bf 11}, 482-486 (2015).

\bibitem{Shekhar-2015} Shekhar, C. {\it et al.} Extremely large magnetoresistance and ultrahigh mobility in the topological Weyl semimetal candidate NbP. {\it Nat. Phys.} {\bf 11}, 645-649 (2015).

\bibitem{Qu-2010}  Qu, D. -X. {\it et al.} Quantum oscillations and Hall anomaly of surface States in the topological insulator Bi$_2$Te$_3$. {\it Science} {\bf 329}, 821-824 (2010).

\bibitem{Wang-2012} Wang, X. {\it et al.} Room Temperature giant and linear magnetoresistance in topological Insulator Bi$_2$Te$_3$ Nanosheets. {\it Phys. Rev. Lett.} {\bf 108}, 266806:1-5 (2012).

\bibitem{Gerhardts-1989} Gerhardts, R. R, Weiss, D. \& Klitzing, K. v. Novel magnetoresistance oscillations in a periodically modulated two-dimensional electron gas. {\it Phys. Rev. Lett.} {\bf 62}, 1173-1176 (1989). 

\bibitem{Mamani-2009} Mamani, N. C. {\it et al.} Nonlinear transport and oscillating magnetoresistance in double quantum wells. {\it Phys. Rev. B} {\bf 80}, 075308:1-8 (2009). 

\bibitem{Tayari-2015} Tayari, V. {\it et al.} Two-dimensional magnetotransport in a black phosphorus naked quantum well. {\it Nat. Commun.} {\bf 6}, 7702:1-6 (2009). 

\bibitem{Wang-2009} Wang, J. {\it et al.} Anomalous magnetoresistance oscillations and enhanced superconductivity in single-crystal Pb nanobelts. {\it Appl. Phys. Lett.} {\bf 92}, 233119:1-3 (2009).

\bibitem{Johansson-2005} Johansson, A. {\it et al.} Nanowire acting as a superconducting quantum interference device. {\it Phys. Rev. Lett.} {\bf 95}, 116805:1-4 (2005). 

\bibitem{Patel-2009} Patel, U. {\it et al.} Magnetoresistance oscillations in superconducting granular niobium nitride nanowires. {\it Phys. Rev. B} {\bf 80}, 012504:1-4 (2009). 

\bibitem{Sochnikov-2010} Sochnikov, I. {\it et al.} Large oscillations of the magnetoresistance in nanopatterned high-temperature superconducting films. {\it Nature Nanotech.} {\bf 5}, 516-519 (2010). 

\bibitem{Kitada-2011} Kitada, A. {\it et al.} Highly reduced anatase TiO$_{2-\delta}$ thin films obtained via low-temperature reduction. {\it Appl. Phys. Express} {\bf 4}, 035801:1-3 (2011).

\bibitem{Mazumdar-1996} Mazumdar, C. {\it et al.} Anomalous magnetoresistance behavior of $R_2$Ni$_3$Si$_5$ ($R$ = Pr, Dy, Ho). {\it Phys. Rev. B} {\bf 54}, 6069-6072 (1996).

\bibitem{Gardner-1972} Gardner, W. E. {\it et al.} The magnetic properties of rare earth-Pd$_3$ phases. {\it J. Phys. F: Metal Phys.} {\bf 2}, 133-150 (1972).

\bibitem{Pandey-2009a} Pandey, A. {\it et al.} Magnetism in ordered metallic perovskite compound GdPd$_3$B$_x$C$_{1-x}$. {\it J. Magn. Magn. Mater.} {\bf 321}, 2311-2317 (2009).

\bibitem{Johnston-2012} Johnston, D. C. Magnetic susceptibility of collinear and noncollinear Heisenberg antiferromagnets. {\it Phys. Rev. Lett.} {\bf 109}, 077201:1-5 (2012). 

\bibitem{Johnston-2015} Johnston, D. C. Unified molecular field theory for collinear and noncollinear Heisenberg antiferromagnets. {\it Phys. Rev. B} {\bf 91}, 064427:1-27 (2015). 

\bibitem{Carvajal-1993}  Rodr\'iguez-Carvajal, J. Recent advances in magnetic structure determination by neutron powder diffraction. {\it Physica B} 192, 55-69 (1993); see also www.ill.eu/sites/fullprof/

\bibitem{Pandey-2008} Pandey, A. {\it et al.} Negative temperature coefficient of resistance in a crystalline compound. {\it Europhys. Lett.} {\bf 84}, 47007:1-6 (2009).

\bibitem{Brown-2006} Brown, P. J. \& Chatterji, T. Neutron diffraction and polarimetric study of the magnetic and crystal structures of HoMnO$_3$ and YMnO$_3$. {\it J. Phys.: Condens. Matter} {\bf 18}, 10085-10096 (2006).

\bibitem{Kadowaki-1995} Kadowaki, H., Takei, H. \& Motoya, K. Double-Q 120 degrees structure in the Heisenberg antiferromagnet on rhombohedrally stacked triangular lattice LiCrO$_2$. {\it J. Phys.: Condens. Matter} {\bf 7}, 6869-6884 (2006).

\bibitem{Hirakawa-1983} Hirakawa, K. {\it et al.} Magnetic susceptibilities of the frustrated triangular lattice antiferromagnets CsVCl$_3$ and V$X_2$ ($X$=Cl, Br and I): appearance of magnetic liquid in the ordered state. {\it J. Phys. Soc. Jpn.} {\bf 52}, 2882-2887 (2006).

\bibitem{Cho-2005} Cho, B. K. {\it et al.} Anomalous magnetoresistance at low temperatures ($T \le 10$~K) in a single crystal of GdB$_4$. {\it J. Appl. Phys.} {\bf 97}, 10A923:1-3 (2005).

\bibitem{Pandey-2009b} Pandey, A., Mazumdar, C \& Ranganathan, R. Magnetic behavior of binary intermetallic compound YPd$_3$. {\it J. Alloys Compd.} {\bf 476}, 14-18 (2009).

\bibitem{Goetsch-2012} Goetsch, R. J. {\it et al.} Structural, thermal, magnetic, and electronic transport properties of the LaNi$_2$(Ge$_{1-x}$P$_x$)$_2$ system. {\it Phys. Rev. B} {\bf 85}, 054517:1-20 (2012). 

\bibitem{Kittel-2005} Kittel, C. {\it Introduction to Solid State Physics}, 8th ed. (Wiley, New York, 2005). 

\bibitem{Jensen-1991} Jensen, J. \& Mackintosh, A. R. {\it Rare Earth Magnetism: Structures and Excitations}, 1st ed. (Clarendon Press, Oxford, 1991).

\bibitem{Elliott-1964} Elliott, R. J. \& Wedgwood, F. A. The temperature dependence of magnetic ordering in the heavy rare earth metals. {\it Proc. Phys. Soc.} {\bf 84}, 63-75 (1964).

\bibitem{Ellerby-1998} Ellerby, M., McEwen, K. A. \& Jensen, J. Magnetoresistance and magnetization study of thulium. {\it Phys. Rev. B} {\bf 57}, 8416-8423 (1998).

\bibitem{Park-2005} Park, T. {\it et al.} Pressure-tuned first-order phase transition and accompanying resistivity anomaly in CeZn$_{1-\delta}$Sb$_2$. {\it Phys. Rev. B} {\bf 72}, 060410(R):1-4 (2005).

\bibitem{Wilding-1965} Wilding, M. D. \& Lee E. W. Superzone boundary effects in the electrical resistivity of dysprosium. {\it Proc. Phys. Soc.} {\bf 85}, 955-961 (1965).

\bibitem{Mackintosh-1962} Mackintosh, A. D. Magnetic ordering and the electronic structure of rare-earth metals {\it Phys. Rev. Lett.} {\bf 9}, 90-93 (1962).

\bibitem{Yamada-1973} Yamada, H. \& Takada, S.  Magnetoresistance of Antiferromagnetic Metals Due to s-d Interaction. {\it J. Phys. Soc. Jpn.} {\bf 34}, 51-57 (1973).

\bibitem{Sanchez-1976} Sanchez, J. P.  {\it et al.} Electronic and magnetic properties of rare-earth-Sn$_3$ compounds from $^{119}$Sn Mossbauer spectroscopy. {\it J. Phys. C: Solid State Phys.} {\bf 9}, 2207-2215 (1976).

\bibitem{Johnston-2016} Johnston, D. C. Magnetic dipole interactions in crystals. {\it Phys. Rev. B} {\bf 93}, 014421:1-33 (2016).

\bibitem{Zhang-2015} Zhang, C-L. {\it et al.} Signatures of the Adler–Bell–Jackiw chiral anomaly in a Weyl fermion semimetal. {\it Nat. Commun.} {\bf 7}, 10735:1-9 (2015)

\bibitem{Huang-2016} Huang, X. {\it et al.} Observation of the chiral-anomaly-induced negative magnetoresistance in 3D Weyl semimetal TaAs. {\it Phys. Rev. X} {\bf 5}, 031023:1-9 (2016)

\end{thebibliography}
\end{document}